\documentclass[prl,aps,english,superscriptaddress,showpacs,twocolumn,amsmath,amssymb]{revtex4-1}
\usepackage{bm,enumerate,amssymb,amsmath,bbold}
\usepackage{natbib}
\usepackage[normalem]{ulem}
\usepackage{graphicx}
\usepackage{ulem,siunitx}
\usepackage{color}

\definecolor{darkgreen}{rgb}{0,0.6,0}
\definecolor{orange2}{rgb}{0.8,0.8,0.2}
\definecolor{grey}{rgb}{0.6,0.6,0.6}
\definecolor{darkblue}{rgb}{0,0,0.7}

\usepackage{pgf}
\usepackage{subfigure}

\usepackage{hyperref}
\usepackage[all]{hypcap}

\usepackage{pgfplots}
\pgfplotsset{compat=newest}
\pgfplotsset{plot coordinates/math parser=false}
\usetikzlibrary{plotmarks}
\usetikzlibrary{calc}

\newcommand{\figdir}{.}

\renewcommand{\emph}{\textit}
\date{November 6, 2014}

\begin{document}

\title{Gapless chiral spin liquid in a kagome Heisenberg model}

\author{Samuel Bieri}

\email{samuel.bieri@alumni.epfl.ch}

\author{Laura Messio}

\author{Bernard Bernu}

\author{Claire Lhuillier}


\affiliation{Laboratoire de Physique Th\'eorique de la Mati\`ere Condens\'ee, Universit\'e Pierre et Marie Curie - Paris VI, CNRS UMR 7600, F-75252 Paris, France}

\begin{abstract}
Motivated by recent experiments on the Heisenberg $S=1/2$ quantum spin liquid candidate material kapellasite, we classify all possible chiral (time-reversal symmetry breaking) spin liquids with fermionic spinons on the kagome lattice. We obtain the phase diagram for the physically relevant extended Heisenberg model, comparing the energies of a wide range of microscopic variational wave functions. We propose that, at low temperature, kapellasite exhibits a gapless chiral spin liquid phase with spinon Fermi surfaces. This two-dimensional state inherits many properties of the nearby one-dimensional phase of decoupled antiferromagnetic spin chains, but also shows some remarkable differences. We discuss the spin structure factors and other physical properties.
\end{abstract}


\maketitle

When low dimensionality, geometric frustration, and antiferromagnetism conspire in quantum magnets, completely novel and exotic physics can emerge at low temperature. In such quantum spin liquid (QSL) phases, the picture of classical magnetic moments breaks down, and fractionalized spinon quasiparticles with unusual statistics and long-range entanglement properties become relevant~\cite{Balents10_Nature_464_199, *Lee08_Science_321_5894}. After intense theoretical research activity on quantum spin liquids in the late 1980's and 1990's due to their intimate relation with high-temperature superconductivity \cite{Lee_RMP_78_2006}, interest in QSL has recently regained momentum because of possible applications in quantum computing~\cite{Kitaev20062}. More strikingly, however, enormous experimental progress in synthetization and characterization of actual spin liquid candidate materials has catapulted the field to an unprecedented stage of maturity in this century (see Refs.~\cite{coldea2001a, Yamashita2008, Itou2010, haidong_cu, haidong_ni, Zorko2008, Zhou2012, Han2012, Helton07, Mendels2007, Colman2008, *Colman2010, Kermarrec14_PRB.90.205103, Fak12_PRL_109_037208} and references therein).

A highly interesting, recently synthesized QSL candidate material is the so-called kapellasite~\cite{Colman2008,*Colman2010, Fak12_PRL_109_037208, Bernu_PRB_87_155107, Kermarrec14_PRB.90.205103}: X-ray diffraction on powder samples of this strong Mott insulator indicates geometrically perfect, uncoupled two-dimensional kagome layers of spin $S=1/2$ Cu ions, despite some on-site Cu/Zn dilution. Muon spectroscopy shows the absence of frozen moments, inelastic neutron scattering exhibits a continuum of excitations (mimicking a spinon continuum), and the plateau in $1/T_1$ of NMR measurements confirms a fluctuating behavior down to 20~mK. 
Experiments on kapellasite therefore provide quite strong evidence in favor of a genuine gapless QSL phase in this material.

In contrast to its polymorph herbertsmithite~\cite{Helton07, Mendels2007} -- one of the best studied QSL candidate to date -- kapellasite is known to have important exchange interactions between farther-neighbor sites in the kagome plane~\cite{Janson_2008, Valenti_PRB.88.075106}. Recent accurate high-temperature series expansions and their fits to susceptibility and specific heat data revealed ferromagnetic interactions on first and second neighbors, while a dominant antiferromagnetic exchange of $J_d\simeq16$~K is present across the hexagons of the lattice \cite{Bernu_PRB_87_155107, Fak12_PRL_109_037208}. These results open up exciting new theoretical prospects, because {\it classical} spin models on the kagome lattice with such farther-neighbor interactions are known to exhibit {\it nonplanar} N\'eel phases, i.e., a spontaneous breaking of time-reversal symmetry (TRS)~\cite{Messio11_PRB_83_184401}. Whether such chiral symmetry breaking can carry over to the quantum regime in a spin liquid phase is one of the central questions we want to address in this paper. Ideas for chiral spin liquids (CSLs) were presented some time ago \cite{KalmeyerLaughlin87_PRL_59_2095, *KalmeyerLaughlin89_PRB.39.11879, WWZ89_PRB_39_11413, Yang93_PRL.70.2641}, but despite the initial excitement and intense research efforts, the lack of realistic theoretical models and experimental realizations of such exotic phases finally led to a stagnation in activity. Recently, however, density matrix renormalization group~(DMRG) computations on a farther-neighbor kagome antiferromagnet found evidence for a gapped CSL state~\cite{Gong2014}.

\begin{figure}[h]
\includegraphics[width=.93\columnwidth]{\figdir/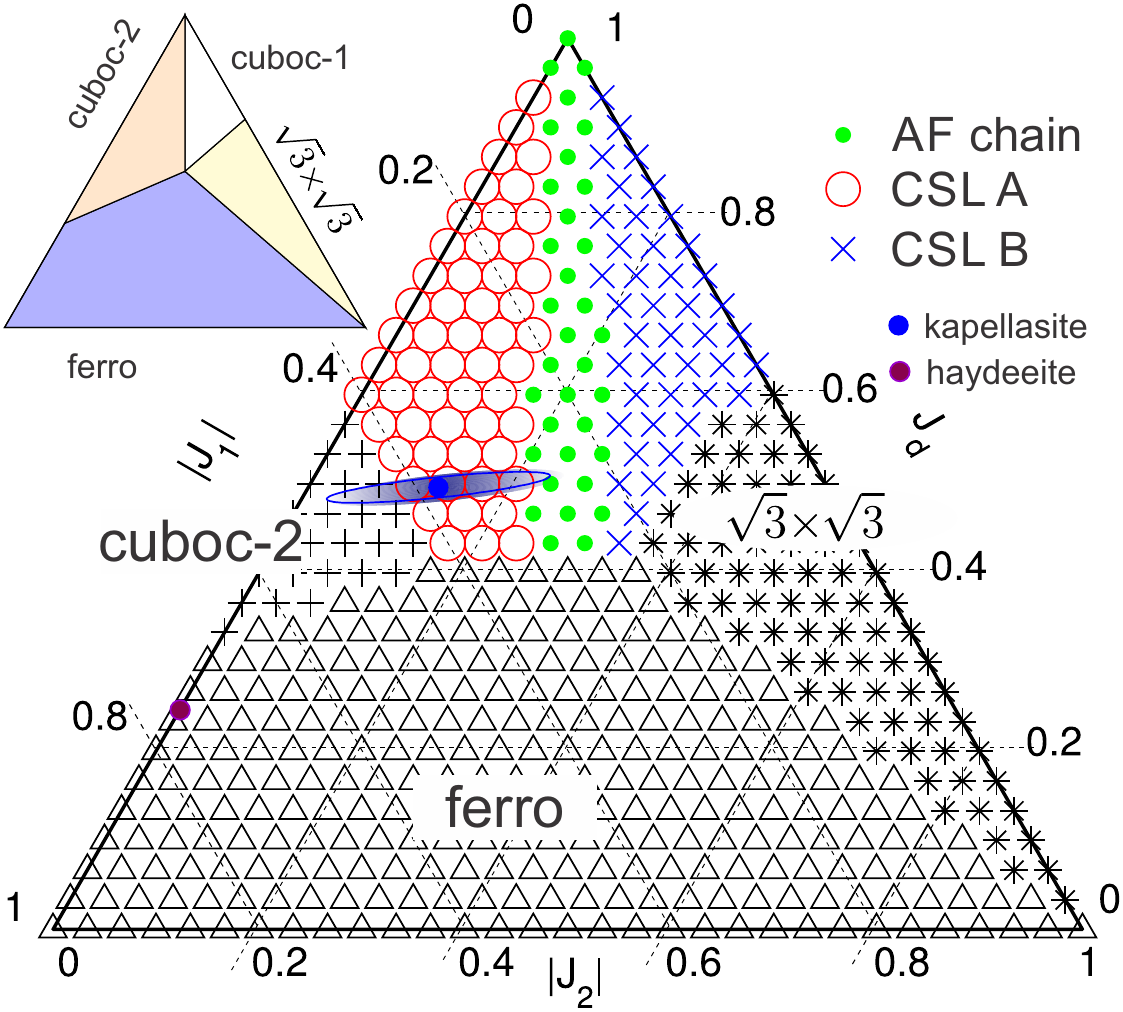}
\caption{
(Color online) Quantum phase diagram of model (\ref{eq:model}) for $J_d\geq0$ and $J_1, J_2\leq0$, with $|J_1|+|J_2|+J_d = 1$. Horizontal lines are $J_d = $~cst. The blue area represents parameters best describing kapellasite~\cite{Bernu_PRB_87_155107}, and the purple dot is the haydeeite kagome ferromagnet~\cite{Boldrin_PRB.91.220408}. Left inset: Classical phase diagram.
\label{fig:phase_diag}}
\end{figure}

In this paper we present a phase diagram of the extended quantum Heisenberg model on the kagome lattice,
\begin{equation}\label{eq:model}
H = J_1\sum_{\langle i,j\rangle} \mathbf{S}_i \cdot \mathbf{S}_j +
J_2 \sum_{\langle\langle i,j\rangle\rangle} {\bf S}_i \cdot {\bf S}_j
+ J_d \sum_{\langle i,j\rangle_d} {\bf S}_i \cdot {\bf S}_j
\end{equation}
with ferromagnetic interactions on first and second neighbors ($J_1\leq0$, $J_2\leq0$), and antiferromagnetic interactions $J_d\geq0$ across the diagonals of the hexagons~\cite{JScale}. Testing a wide range of microscopic spin liquid and correlated N\'eel wave functions, we find the variational phase diagram presented in Fig.~\ref{fig:phase_diag}. In the inset, the phase diagram for the corresponding classical spin model is displayed. The phase dubbed $\sqrt3\times\!\sqrt 3$ has a planar order and a nine-site cell, while \mbox{cuboc-1} \mbox{and cuboc-2} have nonplanar spins in a twelve-site cell~\cite{Messio11_PRB_83_184401}. For $J_d\gtrsim 0.5$, we find three types of QSL phases: a dimensionally reduced, quasi-one-dimensional~(1D), and two TRS breaking phases. We elaborate on these results in the remainder of the paper.

{\it N\'eel ordered states.} Due to the presence of ferromagnetic couplings which favor local transient creation of spin $>1/2$, it can be suspected that semi-classical N\'eel phases can survive in large parts of the phase diagram. Black symbols in Fig.~\ref{fig:phase_diag} designate points where long-range ordered states have the lowest energy. We obtain these wave functions by the Huse-Elser construction \cite{Huse88_PRL_60_2531, Trumper99_PRL.82.3899},
\begin{equation}\label{eq:neel}
  |\text{Neel}\rangle = e^{-\sum_{i,j}{\mathcal J}_{i j} S^z_i S^z_j } \prod_j |{\bm S}\rangle_j\,,
\end{equation}
i.e., nontrivial quantum fluctuation is introduced in product states of classical spin orders by \mbox{first-,} \mbox{second-,} and diagonal-neighbor Jastrow factors. This class of wave functions is known to give excellent variational estimates of the ground-state energy for ordered quantum spin systems~\cite{Huse88_PRL_60_2531, Trumper99_PRL.82.3899}. Because of the presence of ferromagnetic interactions on first- and second-neighbor links, we also checked U(1) liquids (see below) with {\it spin-triplet} fields on these links. However, these spin-rotation broken phases are never stabilized against the singlet liquids anywhere in our phase diagram \cite{tripletHopping}. Nevertheless, further studies of a possible competition of N\'eel states with nematic liquids could be interesting~\cite{Momoi09_PRB.80.064410, Momoi11_PRB.84.134414, Bhatta12_PRB.85.224428, Reuther14_PRB.90.174417}.


{\it QSL wave functions.} Since the putative spin liquid in kapellasite has gapless excitations, we choose to fractionalize spin into fermionic spinon operators $(f_\alpha) = (f_\uparrow, f_\downarrow)$ as $2 S^a = f_\alpha^\dagger\sigma_a^{\alpha\beta}f_\beta$, where $\sigma_a$ are Pauli matrices.
This spin representation introduces a gauge redundancy ${\bm \psi}=(f_\uparrow,f_\downarrow^\dagger)^T\mapsto g {\bm \psi}$, where $g$ is any SU(2) matrix. The state is then constructed via a quadratic spinon Hamiltonian
\begin{equation}\label{eq:Hmf}
  H_0 = \sum_{ij}\xi_{ij} f_{i\alpha}^\dagger f_{j\alpha} + \Delta_{ij} [f_{i\uparrow} f_{j\downarrow}-f_{i\downarrow} f_{j\uparrow}] + \text{h.c.}\,,
\end{equation}
and the microscopic wave function is its Gutzwiller-projected ground state $|\text{QSL}\rangle = \prod_j n_j(2-n_j)|\psi_0\rangle$ \cite{Gros89_AnnPhys_189_53, Bieri12_PRB_86_224409, pathIntMF}. Note that for chiral spin liquids, the parameters $\xi_{ij}$ and $\Delta_{ij}$ are complex in general. At the level of the effective mean-field theory, the SU(2) gauge redundancy is generically broken down to U(1) or $\mathbb{Z}_2$. The theoretical challenge is to exhaustively enumerate all possible liquid phases of the form (\ref{eq:Hmf}) that follow certain symmetry requirements. For this we use the projective symmetry group (PSG) approach introduced by Wen~\cite{Wen02_PRB.65.165113}, and subsequently applied to the kagome lattice in Refs.~\cite{LuRanLee11_PRB.83.224413, YBK13_PRB.88.224413}. In this paper, we significantly extend previous results by systematically classifying time-reversal symmetry broken fermionic QSL phases on the kagome lattice. We are interested in phases with unbroken translation and spin rotation symmetries. The CPT theorem implies that breaking of time reversal is accompanied by breaking of a reflection symmetry. The kagome lattice has a $\pi/3$ rotation symmetry $R$, and reflection symmetries $\sigma$ and $\sigma'$, related by $R\sigma = \sigma'$. We therefore find three possible ways how TRS can break on this lattice: (a) $R$ is intact, and all reflection symmetries $\sigma$ and $\sigma'$ are broken; (b) $R$ and $\sigma$ are broken, and $\sigma'$ is intact; (c) $R$ and $\sigma'$ are broken, and $\sigma$ is intact~\cite{SupMat}.
We label these three types of chiral symmetry breaking by $(\tau_\sigma,\tau_R) = (1,0)$, $(1,1)$, and $(0,1)$, respectively. The fourth case $(\tau_\sigma,\tau_R)=(0,0)$ corresponds to ``symmetric'' spin liquids, i.e., unbroken TRS~\cite{LuRanLee11_PRB.83.224413}. Classical spin states with these symmetries are (a) octahedral state, (b) \mbox{cuboc-1}, and (c) \mbox{cuboc-2}~\cite{Messio11_PRB_83_184401}.

\begin{figure}
\includegraphics[width=1.01\columnwidth]{\figdir/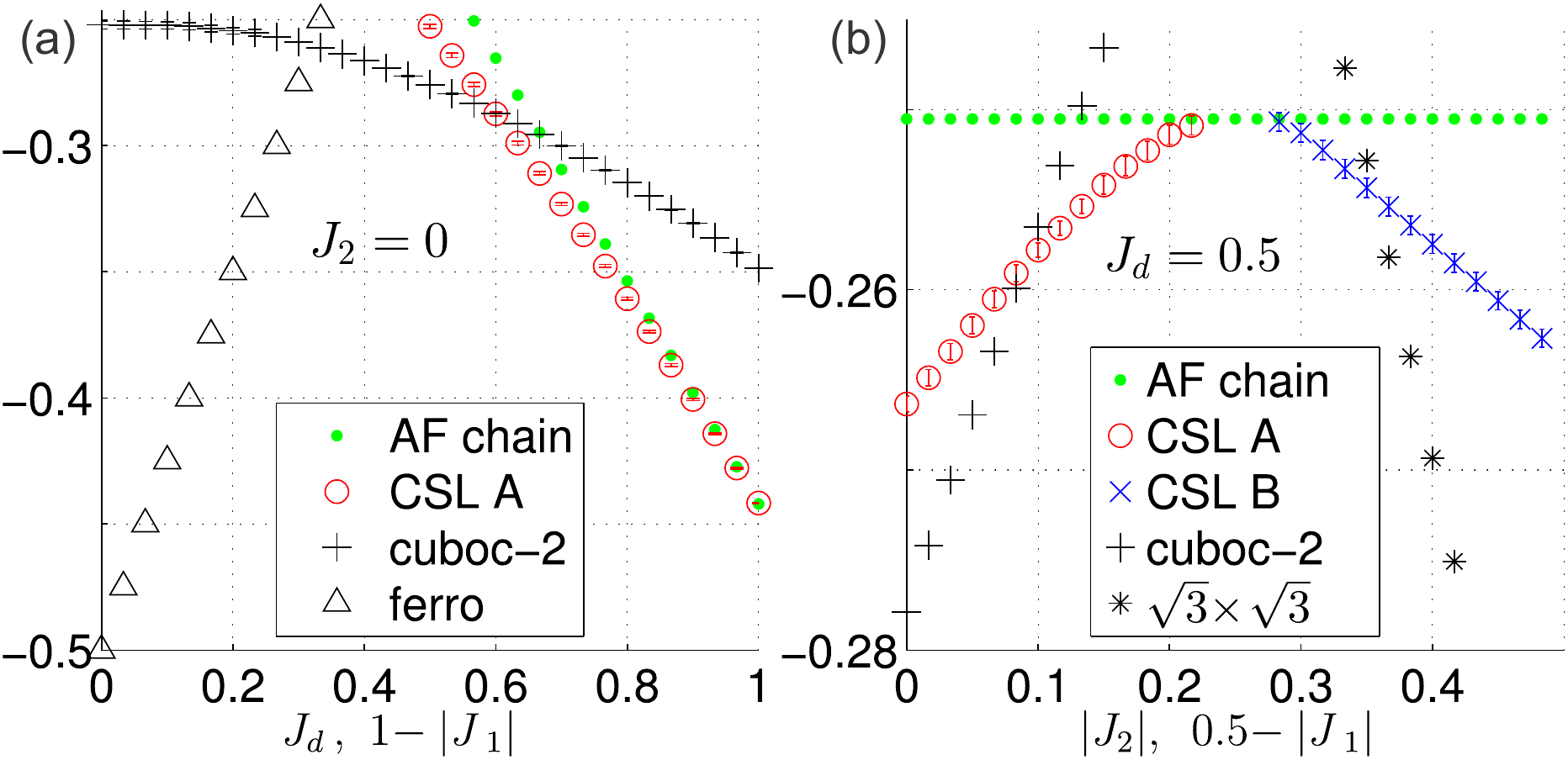}
\caption{
(Color online) Variational energy of microscopic wave functions on the lines (a)~$J_2=0$ and (b)~$J_d=0.5$ of the phase diagram.
\label{fig:var_energy}}
\end{figure}

The algebraic PSG is a projective representation of the lattice symmetries in the SU(2) gauge group~\cite{Wen02_PRB.65.165113}. We find that this representation is not affected by the type of TRS breaking discussed above, and thus there remain 20 gauge inequivalent classes of algebraic PSGs~\cite{LuRanLee11_PRB.83.224413} in the chiral case. However, at the level of mean fields \{$\xi_{ij}$, $\Delta_{ij}$\} compatible with the symmetries, time-reversal plays a crucial role and introduces strong constraints on their possible values. Here, we focus on phases where nonzero mean fields are allowed on {\it at least two} out of first, second, and diagonal links of the kagome lattice. Mean fields on all other links are set to zero. This choice is motivated by the spin model~(\ref{eq:model}) we want to study. Given these restrictions, we find in total 25 distinct chiral mean-field phases (\ref{eq:Hmf}) with a $\mathbb{Z}_2$ gauge structure. We leave a detailed investigation of these Z$_2$ QSLs for future work. Instead, we focus on phases where the mean-field gauge group is U(1). We find 15 such U(1) QSL phases, listed in Table~\ref{tab:U1}.
\begin{table}
\begin{tabular}{c|cc|ccc|ccc|c}
\hline\hline
No.& $\tau_{\sigma}$ & $\tau_{R}$ & $\epsilon_2$ & $g_{\sigma}$& $g_R$ & $\beta_1$ & $\beta_2$ & $\beta_d$ & Description\\
\hline
1  & 0 & 0 & $+$ & $\mathbb{1}_2$ & $\mathbb{1}_2$ & $0$ & $0$ & $0$ & large Fermi surface\\
2  & 0 & 0 & $-$ & $\mathbb{1}_2$ & $\mathbb{1}_2$ & $0$ & $0$ & x & Dirac spect.~\cite{Ran07_PRL.98.117205,*Hermele08_PRB.77.224413}\\
\hline
3  & 1 & 1 & $+$ & $\mathbb{1}_2$ & $\mathbb{1}_2$ & $0$ & $\pi/2$ & x & triangular FS\\
4  & 1 & 1 & $+$ & $i\sigma_2$ & $i\sigma_2$ & $0$ & $\beta_2$ & $0$ & large FS\\
5  & 1 & 1 & $-$ & $\mathbb{1}_2$ & $\mathbb{1}_2$ & $0$ & $\pi/2$ & $0$ & Dirac spectrum\\
6  & 1 & 1 & $-$ & $i\sigma_2$ & $i\sigma_2$ & $0$ & $\beta_2$ & x & FS/Dirac\\
\hline
7  & 0 & 1 & $+$ & $\mathbb{1}_2$ & $\mathbb{1}_2$ & $\pi/2$ & $0$ & $\pi/2$ & triangular FS\\
8  & 0 & 1 & $+$ & $\mathbb{1}_2$ & $i\sigma_2$ & $\beta_1$ & $0$ & $\beta_d$ & large FS\\
9  & 0 & 1 & $-$ & $\mathbb{1}_2$ & $\mathbb{1}_2$ & $\pi/2$ & $0$ & x & kagome FS\\
10 & 0 & 1 & $-$ & $\mathbb{1}_2$ & $i\sigma_2$ & $\beta_1$ & $0$ & x & FS/Dirac\\
11 & 0 & 1 & $-$ & $i\sigma_2$ & $\mathbb{1}_2$ & $\beta_1$ & x & $0$ & Dirac spectrum\\
\hline
12 & 1 & 0 & $+$ & $i\sigma_2$ & $\mathbb{1}_2$ & $\beta_1$ & $\beta_2$ & $0$ & large FS\\
13 & 1 & 0 & $-$ & $i\sigma_2$ & $i\sigma_2$ & $\pi/2$ & $\beta_2$ & $\pi/2$ & CSL~A\\
14 & 1 & 0 & $-$ & $\mathbb{1}_2$ & $i\sigma_2$ & $\beta_1$ & $\pi/2$ & $\beta_d$ & CSL~B\\
15 & 1 & 0 & $-$ & $i\sigma_2$ & $\mathbb{1}_2$ & $\beta_1$ & $\beta_2$ & $\pi/2$ & fully gapped \cite{MarstonZheng91, Hastings00_PRB.63.014413}\\
\hline\hline
\end{tabular}
\caption{
QSL phases with U(1) gauge structure, including symmetric ($\tau_\sigma,\tau_R=0$) and chiral ($\tau_\sigma=1$ or $\tau_R=1$) liquids~\cite{chiralNote}. $\epsilon_2=-1$ indicates doubling of the spinon unit cell. $\beta_a = \text{arg}(\xi_a)$ are the allowed hopping phases in (\ref{eq:Hmf}); ``x'' means $\xi_a=0$. (FS: Fermi surface)
\label{tab:U1}}
\end{table}
The first two columns in this table specify the type of TRS breaking. Thus, phases No.~1 and 2 are symmetric liquids with unbroken TRS. Phases No.~3-15 break time reversal. The next three columns specify the algebraic PSG, i.e., the SU(2) representation of lattice translation, reflection, $g_\sigma$, and $\pi/3$ rotation symmetry, $g_R$. We work in a gauge where the representation of $x$ translation is trivial, $g_x = \mathbb{1}_2$; the $y$ translation is staggered according to the sign $\epsilon_2$, $g_y = (\epsilon_2)^x \mathbb{1}_2$. Similar to the U(1) PSG representation for bosonic spinons~\cite{WangVish06_PRB.74.174423, Messio13_PRB_87_125127, *Messio12_PRL.108.207204, Wang_PRB.82.024419, Li_PRB.76.174406}, we can show that there is always a gauge where the point group representations are independent of sublattice site. Furthermore, for U(1) liquids, $\Delta_{ij}=0$ without loss of generality. Finally, columns $\beta_1$, $\beta_2$, and $\beta_d$ in Table~\ref{tab:U1} contain the allowed complex phases of first, second, and diagonal mean-fields $\xi_a$; ``x'' indicates that hopping must vanish by symmetry, $\xi_a=0$; $\beta_a$ means the phase is unconstrained. If allowed by symmetry, the relative hopping amplitudes, signs, and complex phases are free and will be used as variational parameters~\cite{SupMat}. 
In the last column we give some robust properties of these phases~\cite{chiralNote}, but more details will be published elsewhere.

{\it Quantum phase diagram.} Using large-scale Monte Carlo calculations, we determine the optimal parameters minimizing the energy of (\ref{eq:model}) in the correlated N\'eel (\ref{eq:neel}) and in the chiral U(1) QSL wave functions (\ref{eq:Hmf}). The resulting phase diagram is shown in Fig.~\ref{fig:phase_diag}. The energies of the best variational states on the lines $J_2=0$ and $J_d=0.5$ are given in Fig.~\ref{fig:var_energy}. Black symbols in these figures represent long-range ordered phases~\cite{MC}. As expected, when $J_1$ and $J_2$ are dominant, the system is a ferromagnet. When the strength of $J_d$ increases, it exhibits a first-order transition to semi-classical N\'eel phases, either of nonplanar \mbox{cuboc-2} type or with planar $\sqrt{3}\times\!\sqrt{3}$ order, depending on the relative strengths of $J_1$ and $J_2$. With even larger $J_d$, the ground state moves to true QSL phases (color symbols online). At $J_d=1$, the model consists of decoupled antiferromagnetic (AF) chains. The AF chain is a well-known QSL with gapless Fermi points of deconfined spinons. In our approach, this phase is represented by the so-called ``Gutzwiller-RVB'' wave function~\cite{Shastry88_PRL.60.639, *Haldane88_PRL.60.635} -- a projected chain of itinerant fermions -- which is known to be an extremely good approximation to the true ground state.
Surprisingly, as we decrease $J_d$ while keeping $J_1\simeq J_2$, the 1D phase is very robust and it remains the lowest state. This dimensional reduction can be understood if we picture the chains as AF ordered classical spins: Inter-chain couplings induced by nonzero $J_1 = J_2$ exactly cancel, and two-dimensional~(2D) correlations do not lower the energy. For $|J_1|\gtrsim 2 |J_2|$ and $|J_1|\lesssim |J_2|/2$, however, we find truly two-dimensional chiral spin liquid phases, No.~13 and 14 in Table~\ref{tab:U1}, respectively. In the classical model, $J_1=J_2$ is a line of first-order transition from the chiral \mbox{cuboc-2} to the \mbox{cuboc-1} phase. In the quantum case, our results indicate the existence of an extended intermediate phase with essentially 1D character. This phase may give birth to two-dimensional spin liquids through second-order phase transitions, and the presence of quantum critical lines seems plausible.

\begin{figure*}
\includegraphics[width=\textwidth]{\figdir/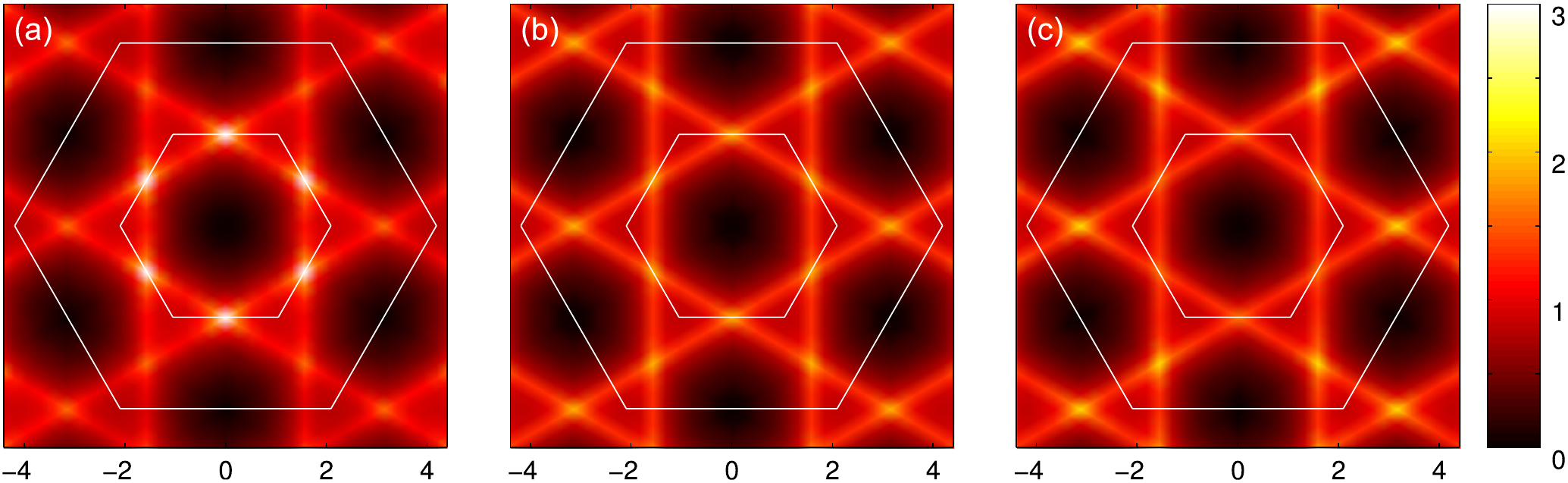}
\caption{(Color online) Static spin structure factor $S({\bm q})$ in the phases (a)~CSL~A, (b)~AF chain, and (c) CSL~B.\label{fig:projSk}}
\end{figure*}

{\it Chiral spin liquids.} Let us discuss the nature of the found 2D QSL states in more detail. We denote the phase for dominant $J_1$ by CSL~A, and the one for dominant $J_2$ by CSL~B. Both are gapless with spinon Fermi surfaces, and they break reflection and TRS, while all other model symmetries are intact. For definiteness, we give the optimal variational parameters at $J = (|J_1|,|J_2|,J_d) = (0.4,0.1,0.5)$ for CSL~A, and at $J = (0.1,0.3,0.6)$ for CSL~B. The hopping amplitudes for CSL~A are $\xi = (|\xi_1|, |\xi_2|, |\xi_d|) \simeq (0.2, 2.7, 7.1)$; the complex hopping phases on first-neighbor and diagonal links are fixed to $\pi/2$ by symmetry (see No.~13 in Table~\ref{tab:U1}), and we find $\beta_2\simeq \pi/5$ for the optimal phase on the second-neighbor link. For CSL~B, we find $\xi\simeq (2.4, 0.2, 7.4)$. Here, the complex phases on first and diagonal links are free (see No.~14), and we find the optimal values $\beta_1\simeq \pi/5$ and $\beta_d\simeq -\pi/25$. QSLs are conveniently characterized by gauge invariant fluxes piercing loops of lattice links~\cite{Wen02_PRB.65.165113, Iqbal11_PRB.83.100404, Clark13_PRL.111.187205}. Using the PSG in Table~\ref{tab:U1}, CSL~A thus has a U(1) flux $\simeq 2\pi/5$ through triangles of second-neighbor links. On the other hand, CSL~B exhibits a flux $\simeq 2\pi/5$ through small lattice triangles and no flux pierces the hexagons.
Note that the optimal hopping phases (and hence the fluxes) on second and first neighbors are simply exchanged in CSL~A and CSL~B~\cite{optParameters}.

{\it Static structure factor.} In Fig.~\ref{fig:projSk}, we display the static spin structure factor $S({\bm q})$ for the three QSL phases calculated on square clusters of $N=3(12)^2$ sites~\cite{SupMat}. In the 1D phase for $J_1\simeq J_2$, the spinon can only propagate along the diagonals of the hexagons, which form arrays of uncoupled chains in three directions $\hat{\bm e}_1$, $\hat{\bm e}_2$, and $\hat{\bm e}_3$. The structure factor for these chains is logarithmically divergent in $N$ at momenta connecting the Fermi points, determined by ${\bm q}\cdot \hat{\bm e}_n=\pi/(2a)$; (nearest-neighbor distance $a = 1$). This explains the lines of intensity in the 1D phase [Fig.~\ref{fig:projSk}(b)]. When moving to the 2D CSL phases in Figs.~\ref{fig:projSk}(a) and \ref{fig:projSk}(c), these general geometric features of $S({\bm q})$ remain. However, the scattering becomes more diffuse, and the passage to 2D is accompanied by a transfer of spectral weight. In CSL~A, the intensity is concentrated at the six inner line crossings ($M$ points), while in CSL~B, it is dominantly at the outer crossings. This is a fading reminiscence of soft points in the classical partners of these phases~\cite{Messio11_PRB_83_184401}.

\textit{Physical properties.}
The blue area in Fig.~\ref{fig:phase_diag} depicts the confidence region of model parameters describing kapellasite according to recent experimental fits~\cite{Bernu_PRB_87_155107}. The most likely spin exchange couplings for this material, $J \simeq (0.38,0.13,0.49)\times 32$~K, is located in the CSL~A phase. This result is in excellent agreement with the neutron scattering data which clearly indicate the absence of N\'eel or nematic order and of their Goldstone modes.  The averaged static spin structure factor of CSL~A with its soft maxima at the $M$ points [Fig.~\ref{fig:projSk}(a)] gives a good description of the powder signal. These experimental properties are also consistent with a bosonic description of the underlying spinons~\cite{Fak12_PRL_109_037208}. However, the decisive strength of the present explanation is the existence of a continuum of gapless $S=1$ excitations near the spinon Fermi surface of the theory. Constructing these (projected) excitations explicitly~\cite{Hermele08_PRB.77.224413}, we estimate a spinon bandwidth of $\sim 2.1~\text{meV}$. These features are lacking in a bosonic approach, and they are crucial in explaining the experimental data.



The found gapless CSLs are expected to exhibit longitudinal spin and heat transport~\cite{Lee10_PRL.104.066403}. Furthermore, spontaneous TRS breaking implies corresponding Hall currents, even in the absence of Dzyaloshinskii-Moriya terms or external magnetic fields. Classical simulations suggested that chiral symmetry breaking can survive up to small but finite temperature~\cite{Domenge_PRB.77.172413, *Messio08_PRB.78.054435}. A mean-field calculation for our CSL phases predicts a linear temperature scaling of specific heat and conductivities, but strong spinon interaction due to U(1) gauge fluctuation is likely to change this result~\cite{Motrunich05_PRB_72_045105, *Motrunich06_PRB.73.155115}. Nevertheless, a nonzero Hall angle is robust and would be a strong experimental indication in favor of our theory.

{\it Conclusion.}
In this paper we studied the zero-temperature phase diagram of the quantum Heisenberg model (\ref{eq:model}) on the kagome lattice with ferromagnetic exchanges $J_1$ and $J_2$, and antiferromagnetic $J_d$ across the hexagons. It harbors not only semi-classical phases, but also genuine chiral spin liquids. On this basis, we propose a quantitative explanation for the scattering pattern measured in powder samples of kapellasite, and we predict the structure factor expected for single crystals. From a theoretical perspective, these CSLs appear through frustration of arrays of AF spin chains. It is a rare occurrence of TRS breaking in a pure Heisenberg model without higher-order spin exchange~\cite{Lauchli03_PRB.67.100409, Mishmash13_PRL.111.157203, Nielsen2013}. This opens interesting avenues for other approaches (1D bosonization~\cite{Azaria98_PRB.58.R8881}, DMRG~\cite{YanHuseWhite_03062011, JiangWengSheng08_PRL.101.117203, Depenbrock2012, He2014, Balents_PRB.86.024424, Bauer2014}, DMFT~\cite{DMFT_RMP.68.13}, fRG~\cite{Thomale14_PRB.89.020408}, cluster methods~\cite{FarnellBishopLi_PRB.89.184407, Auerbach11_PRB.87.161118}, refined VMC~\cite{Iqbal11_PRB.84.020407, Iqbal13_PRB.87.060405, *Iqbal14_PRB.89.020407, Imada14_JPSJ.83.093707, Tay11_PRB.84.020404}, etc) exploring this novel type of dimensional crossover from 1D to 2D quantum phases. In fact, compounds with a network of frustrated AF chains are ubiquitous and their neutron diffraction patterns remain \textit{terra incognita}.

\begin{acknowledgments}
We thank P.~Mendels, B.~F{\aa}k, F.~Bert, M.~Serbyn, and P.~A.~Lee for many helpful discussions. The supercomputer ``Mesu'' of UPMC and the computing cluster of LPTMC were used for the numerical results. UPMC belongs to Sorbonne Universit\'es. This work is supported by ANR-12-BS04-0021.
\end{acknowledgments}

\bibliography{claire,kagome_chiral_biblio,notes}

\begin{thebibliography}{84}
\bibitem [{JSc()}]{JScale}%
  \BibitemOpen
  \href@noop {} {}\bibinfo {note} {We choose an energy scale such that
  $|J_1|+|J_2|+J_d = 1$ throughout this paper.}\BibitemShut {Stop}%
\bibitem [{tri()}]{tripletHopping}%
  \BibitemOpen
  \href@noop {} {}\bibinfo {note} {We also consider fermionic U(1) CSL states
  with complex spin-triplet hopping on first and second neighbors, and singlet
  hopping on the diagonal link. These states break SU(2) spin rotation down to
  U(1) around $S_z$. Using the PSG construction, we systematically check all
  such states, including those breaking the point group up to time reversal.
  These triplet (or ``nematic'' \cite{Momoi09_PRB.80.064410,
  Momoi11_PRB.84.134414, Bhatta12_PRB.85.224428, Reuther14_PRB.90.174417})
  states are always higher in energy, except around $J\simeq(0.3,0.3,0.4)$,
  where they are almost degenerate with the ones shown in the phase diagram.
  Further studies, e.g., of $\mathbb{Z}_2$ triplet states, may be interesting
  in this parameter region.}\BibitemShut {Stop}%
\bibitem [{pat()}]{pathIntMF}%
  \BibitemOpen
  \href@noop {} {}\bibinfo {note} {Note that in a path-integral approach, the
  mean-field theory Eq.~(3) is well behaved even in the presence of
  ferromagnetic couplings, as explained in Ref.~\cite{Auerbach91_PRB.43.7800}.
  In our variational approach, we stress the importance of Gutzwiller
  projection, which can induce ferromagnetic
  correlations~\cite{SupMat}.}\BibitemShut {Stop}%
\bibitem [{chi()}]{chiralNote}%
  \BibitemOpen
  \href@noop {} {}\bibinfo {note} {Some of the QSL states in Table~I (albeit
  only for first neighbors) have been discussed previously
  \cite{MarstonZheng91,Hastings00_PRB.63.014413,Ran07_PRL.98.117205}. We ignore
  states with flat bands at the Fermi energy, for which there is no unique
  ground state. Clearly, all states develop Fermi surfaces when $\xi_d$ becomes
  dominant.}\BibitemShut {Stop}%
\bibitem [{MC()}]{MC}%
  \BibitemOpen
  \href@noop {} {}\bibinfo {note} {The QSL states are calculated on square
  clusters of $N=3(12)^2$ sites. Such large systems are necessary especially
  when $|\xi_d|\gg|\xi_1|,|\xi_2|$ and finite-size effects become strong.
  Spinon boundary conditions are periodic in one, and antiperiodic in the other
  lattice direction. The ordered wave functions are mostly calculated on $N=3
  (24)^2$ sites.}\BibitemShut {Stop}%
\bibitem [{opt()}]{optParameters}%
  \BibitemOpen
  \href@noop {} {}\bibinfo {note} {Accurate determination of parameters is
  difficult in Monte Carlo, because energies (and other physical properties)
  weakly depend on them. In some cases we find slightly different fluxes with
  almost degenerate energy. However, TRS breaking appears to be
  robust.}\BibitemShut {Stop}%
\end{thebibliography}
\begin{thebibliography}{15}
\bibitem [{onS()}]{onSiteTerms}%
  \BibitemOpen
  \href@noop {} {}\bibinfo {note} {We note that on-site terms are present in
  our spinon Hamiltonian. If allowed by symmetry, they are consistently
  adjusted to bring the wave function to half filling before Gutzwiller
  projection. For the U(1) CSL~A and CSL~B states of this paper, it turns out
  that the chemical potentials are pinned to zero by symmetries.}\BibitemShut
  {Stop}%
\bibitem [{TR()}]{TR}%
  \BibitemOpen
  \href@noop {} {}\bibinfo {note} {The choice $\Theta(u_{ij}) = -u_{ij}$ for
  the TRS representation is not crucial for main results of this paper.
  Furthermore, we would like to stress that time reveresal does not enter the
  algebraic PSG classification.}\BibitemShut {Stop}%
\end{thebibliography}

\end{document}


\title{Gapless chiral spin liquid in a kagome Heisenberg model}

\author{Samuel Bieri}


\author{Laura Messio}

\author{Bernard Bernu}

\author{Claire Lhuillier}

\affiliation{Laboratoire de Physique Th\'eorique de la Mati\`ere Condens\'ee, Universit\'e Pierre et Marie Curie - Paris VI, CNRS UMR 7600, F-75252 Paris, France}

\maketitle

In this Supplemental Material, we provide technical details and complementary information that were omitted in the main text of the paper.

\subsection{Fermionic spinon mean fields and PSG classification}

For completeness and to facilitate reproduction of results presented in the main text, we summarize here some key elements of the projective symmetry group (PSG) classification~\cite{Wen02_PRB.65.165113}, and we remind a notation that is widely used in the literature.

The complex hopping $\xi_{ij}$ and pairing $\Delta_{ij}$ parameters in the spinon Hamiltonian Eq.~(3) of the main text are conveniently represented in the matrix form
\begin{equation}\label{eq:meanu}
  u_{ij} = \left(\begin{array}{cc}
    -\xi_{ij}^* & \Delta_{ij}\\
    \Delta_{ij}^* & \xi_{ij}
  \end{array}\right)\,.
\end{equation}
The quadratic spinon Hamiltonian can be written using the SU(2)$_\text{gauge}$ doublet ${\bm \psi} = (f_\uparrow, f_\downarrow^\dagger)^T$ as
\begin{equation}\label{eq:h0}
  H_0 = \sum_{i,j} {\bm \psi}_i^\dagger u_{ij} {\bm \psi}_j + h.c.
\end{equation}
For spin-rotation invariant (singlet) liquids, we have $u_{ji} = u_{ij}^\dagger$~\cite{spinRot}. For the sake of brevity, we omit {\it on-site} terms (chemical potential and pairing) in the present discussion, but refer to the literature instead~\cite{Wen02_PRB.65.165113, onSiteTerms}.

Let $g_j$ be a SU(2) matrix on site $j$ of the lattice. A local ``gauge'' transformation ${\bm \psi}_j \mapsto g_j {\bm \psi}_j$ in the spinon Hilbert space leaves the subspace of spin states ($n_j = f_\alpha^\dagger f_\alpha \equiv 1$) invariant. All transformations (space group or time reversal) acting in the spinon Hilbert space can therefore be supplemented by additional SU(2) gauge rotations. The symmetry generators follow (lattice specific) algebraic relations, and their SU(2) gauge representation must be compatible with these relations. The classification of algebraic PSG consists in listing all gauge-inequivalent symmetry representations that are compatible with these algebraic relations.

\subsection{Symmetries of the kagome lattice and their SU(2) representation}

\begin{figure}
  \includegraphics[width=.4\columnwidth]{\figdir/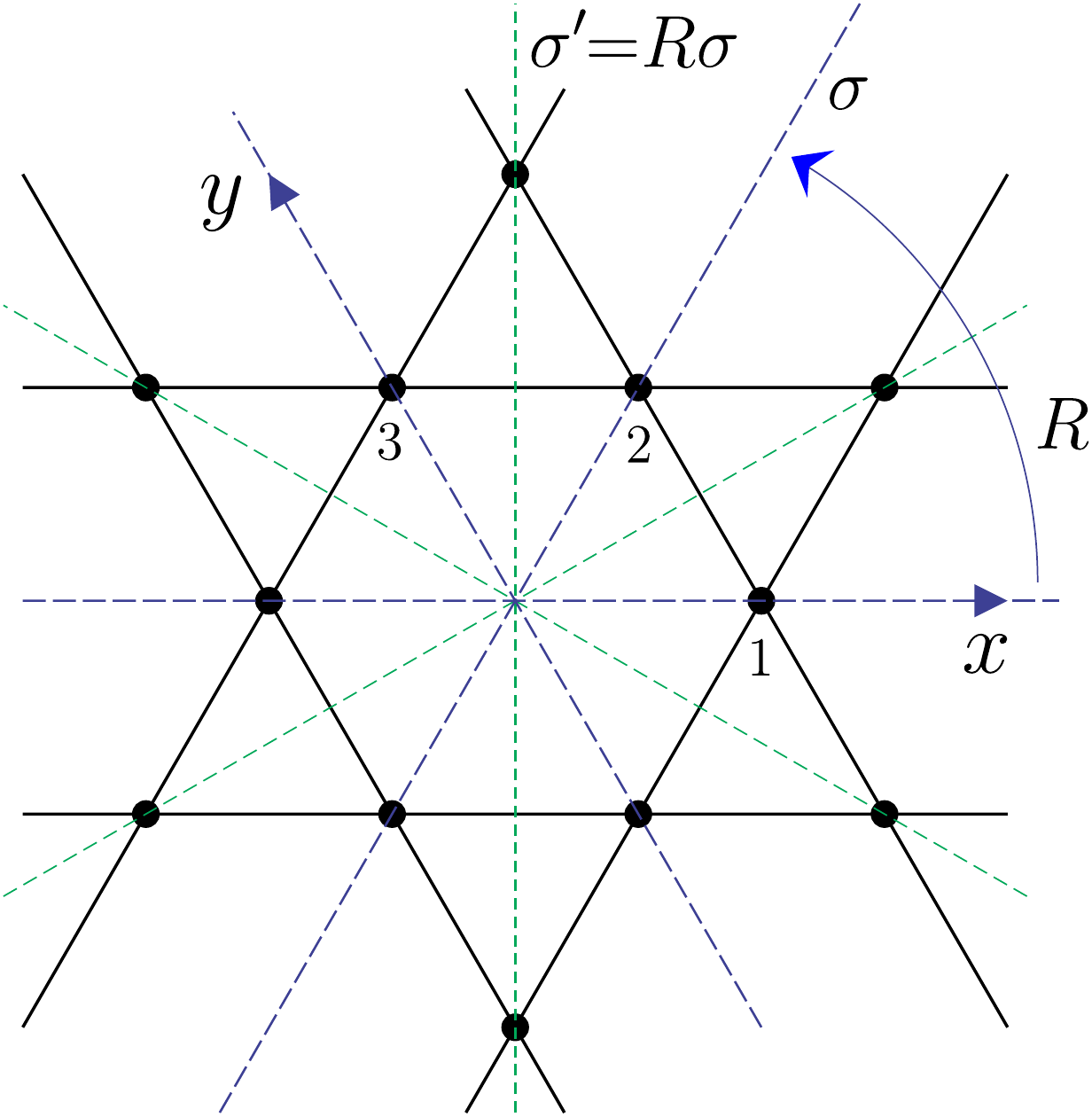}
\caption{
Point group generators $R$ and $\sigma$ of the kagome lattice. The reflection symmetry $\sigma'\! = R\sigma$ is also shown, as well as translations in $x$ and $y$ direction. The sites labelled by $1$, $2$, $3$ are chosen as a unit cell of the lattice.
\label{fig:symmetries}
}
\end{figure}

The point group generators, $\sigma$ (reflection) and $R$ (rotation by $\pi/3$), as well as translations $x$ and $y$ of the kagome lattice, as used in the main text, are illustrated in Fig.~\ref{fig:symmetries}. For the unit cell, we choose the sites $1$, $2$, and $3$ shown in the same figure. The reflection symmetry $\sigma'\! = R\sigma$ used in the main text is also shown.

The PSG representation of lattice symmetries can be written in the form
\begin{subequations}\begin{align}
  g_x(x,y) &= \mathbb{1}_2\,,\label{eq:psg1}\\
  g_y(x,y) &= (\epsilon_2)^x \mathbb{1}_2\,,\label{eq:psg2}\\
  g_\sigma(x,y) &= (\epsilon_2)^{x y} g_{\sigma}\,,\label{eq:psg3}\\
  g_R(x,y) &= (\epsilon_2)^{x y + y(y+1)/2} g_{R}\,,\label{eq:psg4}
\end{align}\end{subequations}
where $x$ and $y$ are the sublattice coordinates shown in Fig.~\ref{fig:symmetries}. The sign $\epsilon_2 = \pm$ and the constant matrices $g_\sigma$ and $g_R$ are given in Tab.~I of the main text for the gauge inequivalent PSG solutions of interest.

The matrix of mean-field parameters (\ref{eq:meanu}) on a given link $(ij)$ may be propagated to another link using symmetry operations. At this point, the PSG representation enters the scene. For example, rotation $R$ acts in the following way on the mean-field,
\begin{equation}
  R( u_{ij} ) = g_R(i) u_{R^{-1}(i) R^{-1}(j)} [g_R(j)]^\dagger\, .
\end{equation}
Since the mean field $u_{ij}$ must respect all lattice symmetries (up to time reversal and gauge transformations), $u_{ij}$ on the rotated link can be identified with $u_{ij}$ on the unrotated link. Time reversal is implemented in the spinon Hilbert space as $\Theta(u_{ij}) = - u_{ij}$~\cite{TR}. For the rotation symmetry $R$, we therefore find
\begin{equation}
  u_{ij} = (-)^{\tau_R} g_R(i) u_{R^{-1}(i) R^{-1}(j)} [g_R(j)]^\dagger\,,
\end{equation}
where $\tau_R\in \{0,1\}$ is the time-reversal parity under rotation in the chiral spin liquid state, as discussed in the main text of the paper. Similarly, the mean field can be propagated by translation, using Eqs.~(\ref{eq:psg1}) and (\ref{eq:psg2}).

\begin{figure}
  \includegraphics[width=.4\columnwidth]{\figdir/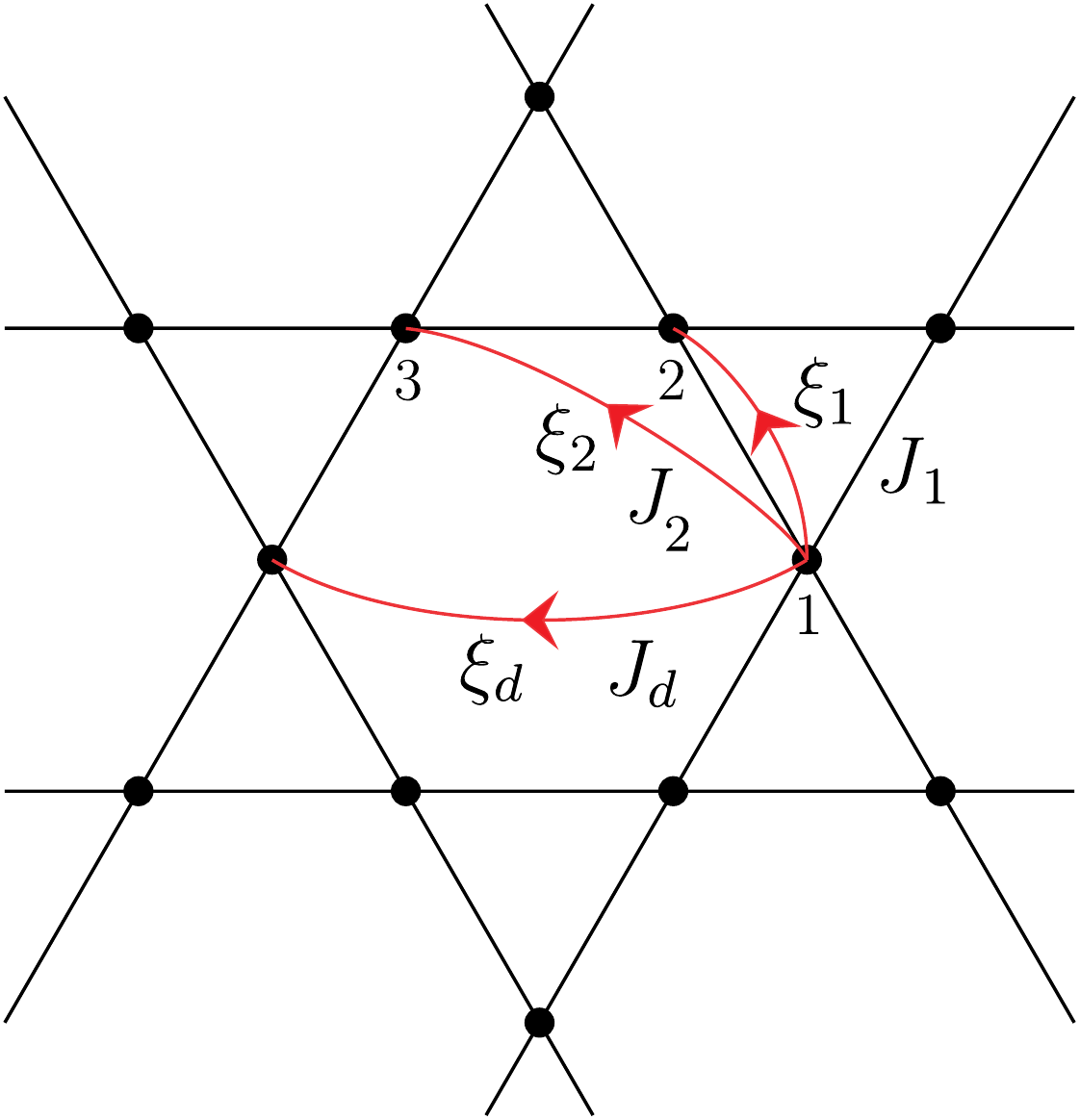}
\caption{
Hopping parameters $\xi_a$ on first, second, and diagonal lattice links as used in the main text. The propagation of these parameters to the entire lattice is then done using the algebraic PSG. The exchange interactions $J_1$, $J_2$, and $J_d$ as used in Eq.~(1) of the main text are also given.
\label{fig:hoppings}
}
\end{figure}

The choice of hopping mean fields on first, second, and diagonal lattice links are displayed in Fig.~\ref{fig:hoppings}. For U(1) liquids, there is always a gauge where the pairing $\Delta_{ij}=0$. The complex hopping phases allowed by the symmetry constraints are given in Tab.~I of the main text. To construct the spinon Hamiltonian on the entire lattice, it is convenient to start by rotating these mean fields around the hexagon using the PSG representation of $R$ as discussed above. Finally, the rotated mean fields can be propagated to the entire lattice by translation in a similar way.

\subsection{Spinon Fermi surface of CSL~A}

\begin{figure}
  \includegraphics[width=.4\columnwidth]{\figdir/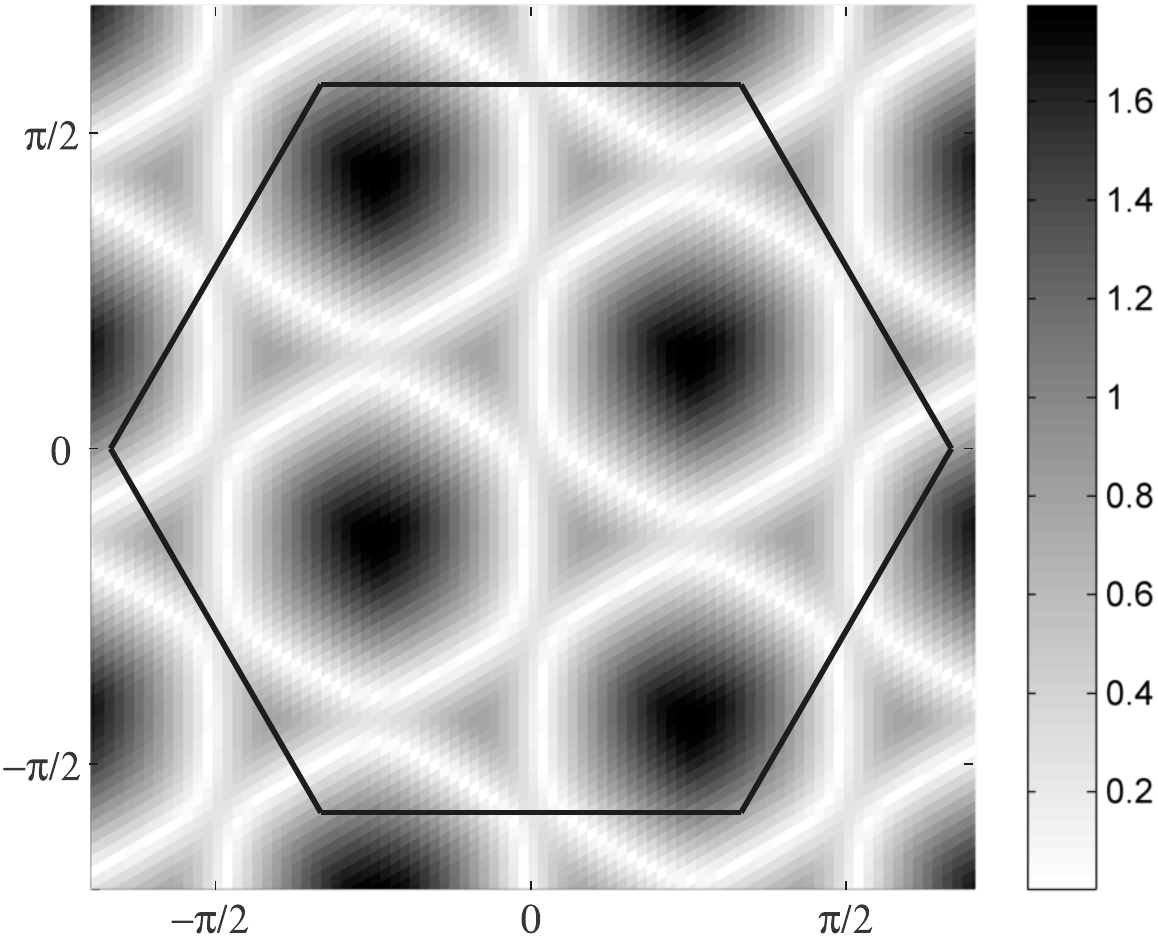}
\caption{
Energy $E$ (in meV) of the lowest spinon mean-field band in CSL~A for the optimal parameters $\xi=(0.2,2.7,7.4)$ and hopping phases $\beta = (0.5,0.2,0.5)\pi$ at $J=(0.4,0.1,0.5)$. The spinon bandwidth is $2.1$~meV. White lines are spinon Fermi surfaces ($E=0$), and the black hexagon is the Brillouin zone.
\label{fig:FS_CSL_A}}
\end{figure}

The lowest spinon mean-field band for CSL~A is displayed in Fig.~\ref{fig:FS_CSL_A}, for the optimal parameters at exchange coupling $J=(0.4,0.1,0.5)$; (describing kapellasite). The color code is such that the white lines represent the spinon Fermi surfaces. The first Brillouin zone of the lattice is shown in black. Note that $\epsilon_2=-1$ in this PSG (no.~13 in Tab.~I of the main text) means that the spinon unit cell is doubled, and it contains 6 lattice sites. This translation (and rotation) symmetry breaking in the spinon spectrum is only a gauge artefact, and all physical quantities (spectral functions, two-spinon excitations, etc) respect those symmetries (see also~\cite{Ran07_PRL.98.117205,*Hermele08_PRB.77.224413}). The mean-field bandwidth in Fig.~\ref{fig:FS_CSL_A} is scaled to the energy cost of a projected spin $S=1$ excitation, going from the Fermi surface to the highest spinon band ($\simeq 2.1$~meV).

\subsection{Gutzwiller projection and ferromagnetism}

In this study, the Gutzwiller projector in the construction of the QSL wave functions is crucial. At the quadratic mean-field level (i.e., without Gutzwiller projection) and for a spin-rotation invariant QSL, one can show that the two-point function is given by
\begin{equation}\label{eq:twopoint}
  \langle {\bm S}_i\cdot {\bm S}_j\rangle_\text{MF} = -\frac{3}{8} ( |\langle\hat\chi_{ij}\rangle|^2 + |\langle\hat\eta_{ij}\rangle|^2 )
\end{equation}
in the standard notation $\hat\chi_{ij} = f_{i\alpha}^\dagger f_{j\alpha}$ and $\hat\eta_{ij} = \varepsilon_{\alpha\beta}f_{i\alpha} f_{j\beta}$~\cite{Wen02_PRB.65.165113}. That is, the spin correlators cannot be ferromagnetic in a mean-field approach of Abrikosov fermions. This fact is to be contrasted with Schwinger boson mean-field theory~\cite{Sachdev92_PRB.45.12377, WangVish06_PRB.74.174423, Messio12_PRL.108.207204, *Messio13_PRB_87_125127}, where hopping gives ferromagnetic correlators, while pairing results in anti-ferromagnetic ones.

The highly accurate approximation of ground state (and low-lying excitations) of the anti-ferromagnetic Heisenberg spin-1/2 chain by fermionic spinons~\cite{Yunoki06_PRB.74.014408, Shastry88_PRL.60.639, *Haldane88_PRL.60.635} is only achieved once the Gutzwiller projector is applied to the Slater determinant. Only then, the staggered nature of the correlation function and the correct long-distance power law are recovered, $S(x) \propto (-)^x/|x|$. It is therefore no surprise that Gutzwiller projection has strong effects on the two-dimensional QSL states discussed in this paper. For instance, in the limit of decoupled arrays of chains, the unphysical inter-chain scattering of mean-field spinons is merely suppressed by the projection. For the two-dimensional CSL phases discussed in this paper, this projection induces ferromagnetic correlators on first or second neighbors of the Kagome lattice, which leads to their energetic competitiveness.

\subsection{Ordered moments}

\begin{figure}
  \includegraphics[width=.6\columnwidth]{\figdir/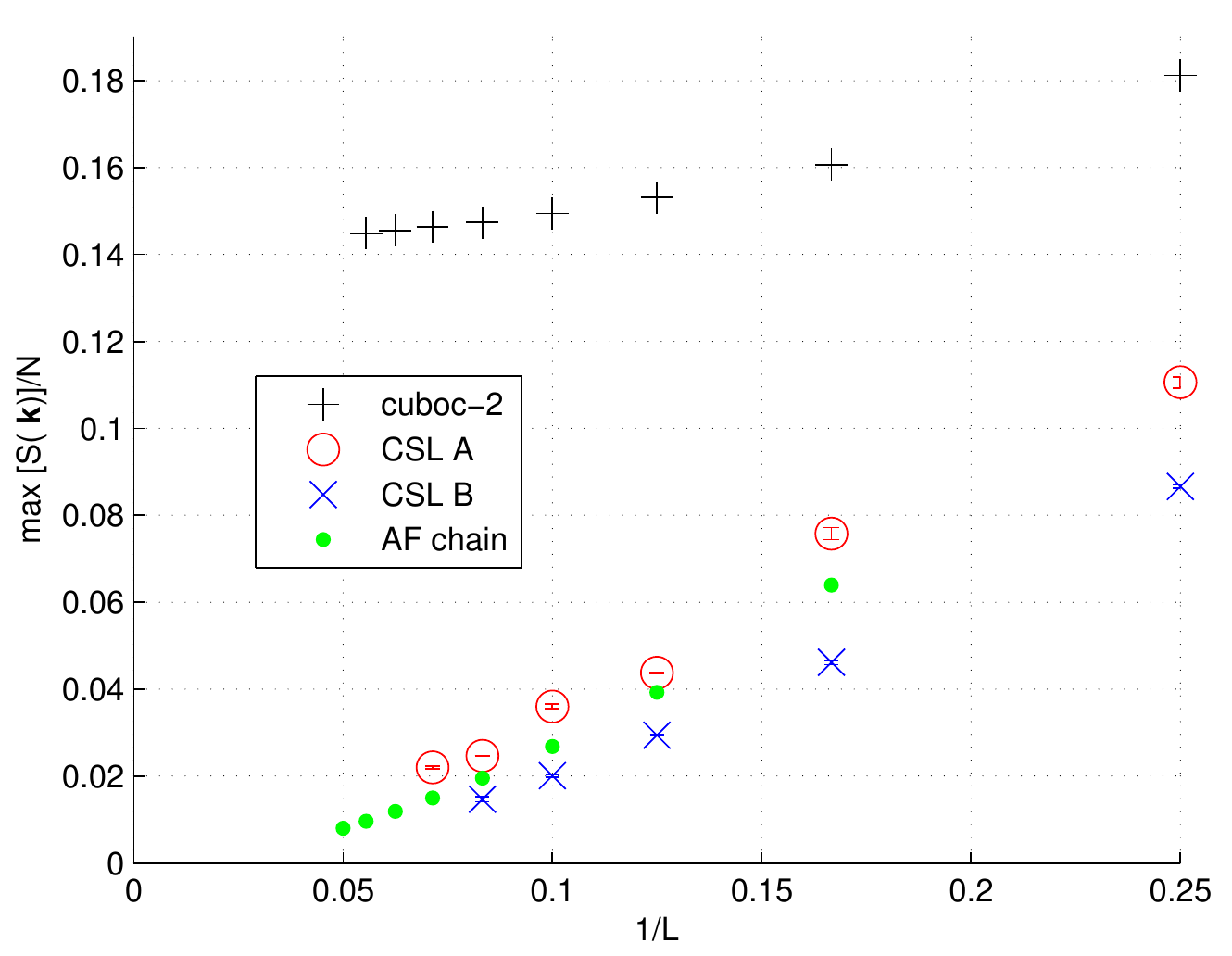}
\caption{
Scaling of the peaks $\text{max}[S(\bm k)]/N$ as a function of $1/L$ for the three QSL phases and for the semi-classical cuboc-2 phase discussed in this paper. The CSL states have fixed variational parameters, as given in the main text; CSL~A: $\xi = (0.2,2.7,7.1)$, $\beta = (0.5,0.2,0.5)$; CSL~B: $\xi = (2.4, 0.2, 7.4)$, $\beta = (0.2,0.5,-0.04)\pi$.
\label{fig:mmts_scaling}}
\end{figure}

The attentive reader may be worried that the pronounced peaks in the spin structure factors $S(\bm k)$ for the QSL phases displayed in Fig.~3 of the main text could indicate a symmetry breaking and long-range ordering as the system size grows. In fact, for the anti-ferromagnetic~(AF) spin chain, it is known that the structure factor at the AF wave vector diverges logarithmically with system size. In Fig.~\ref{fig:mmts_scaling}, we display a finite-size scaling of the ordered moment for the QSL phases discussed in the main text. We conclude from this analysis that the peaks in the 2D CSL phases are at most logarithmically divergent, and these phases do not exhibit translation symmetry breaking.

For comparison, we also display the ordered moment of the semi-classical cuboc-2 state in Fig.~\ref{fig:mmts_scaling} for parameters close to the transition line, $J \simeq (0.4,0.1,0.5)$. The ordered moment of this wave function is clearly finite in the thermodynamic limit, with a magnitude of approximately $80\%$ of the classical value. It should be noted that while the relatively simple Huse-Elser wave functions we use in this study [Eq.~(2) of the main text] provide good estimates of the ground state energy, it is known that the ordered moment is generally overestimated in this approach (see, e.g.,~\cite{LiBishopCampbell15_PRB.91.014426} and refs.\ therein).

\bibliography{claire,kagome_chiral_biblio,notes}